\documentclass{aa}
\usepackage{times,graphicx,url}
\usepackage{natbib}
\usepackage{bm,color}
\def\blue{\textcolor{black}}
\graphicspath{{./fig/}{./png/}}
\newcommand{\EQ}{\begin{equation}}
\newcommand{\EN}{\end{equation}}
\newcommand{\EQA}{\begin{eqnarray}}
\newcommand{\ENA}{\end{eqnarray}}

\newcommand{\Eq}[1]{Eq.~(\ref{#1})}

\newcommand{\Sec}[1]{Sect.~\ref{#1}}

\newcommand{\Fig}[1]{Fig.~\ref{#1}}

\newcommand{\bra}[1]{\langle #1\rangle}

\newcommand{\meanrho}{\overline{\rho}}

{}
{}
\newcommand{\meanFFFF}{\overline{\mbox{\boldmath ${\cal F}$}}{}}{}

\newcommand{\meanSSSS}{\overline{\mbox{\boldmath ${\mathsf S}$}} {}}

\newcommand{\meanSSS}{\overline{\mathsf{S}}}
{}
{}
{}
{}
{}
{}
{}
{}
\newcommand{\meanAA}{\overline{\mbox{\boldmath $A$}}{}}{}
\newcommand{\meanBB}{\overline{\mbox{\boldmath $B$}}{}}{}
{}
{}
{}
{}
{}
{}
{}
{}
\newcommand{\meanJJ}{\overline{\mbox{\boldmath $J$}}{}}{}
{}
\newcommand{\meanUU}{\overline{\bm{U}}}

{}
{}
{}

\newcommand{\meanB}{\overline{B}}

\newcommand{\meanU}{\overline{U}}

\newcommand{\hatk}{\hat{k}}
\newcommand{\hatkk}{\hat{\bm{k}}}

{}

{}
\newcommand{\hatNNN}{\hat{\mbox{\boldmath ${\cal N}$}}{}}{}
\newcommand{\hatNN}{\hat{{\cal N}}}{}
{}

\newcommand{\kk}{\bm{k}}

\newcommand{\uu}{\mbox{\boldmath $u$} {}}
\newcommand{\UU}{\mbox{\boldmath $U$} {}}

\def\bb{\bm{b}}

\newcommand{\BB}{\mbox{\boldmath $B$} {}}

\newcommand{\JJ}{\mbox{\boldmath $J$} {}}

\newcommand{\AAA}{\mbox{\boldmath $A$} {}}

\newcommand{\ff}{\mbox{\boldmath $f$} {}}

\newcommand{\FF}{\mbox{\boldmath $F$} {}}

\newcommand{\grav}{\mbox{\boldmath $g$} {}}
\newcommand{\nab}{\mbox{\boldmath $\nabla$} {}}
\newcommand{\OO}{\bm{\Omega}}
\newcommand{\oo}{\mbox{\boldmath $\omega$} {}}

\newcommand{\SSSS}{\mbox{\boldmath ${\sf S}$} {}}

\newcommand{\ii}{{\rm i}}

\newcommand{\bec}[1]{\mbox{\boldmath $ #1$}}
\newcommand{\DD}{{\rm D} {}}

\def\Beqz{B_{\rm eq0}}

\def\ga{\mathrel{\mathchoice {\vcenter{\offinterlineskip\halign{\hfil
$\displaystyle##$\hfil\cr>\cr\sim\cr}}}
{\vcenter{\offinterlineskip\halign{\hfil$\textstyle##$\hfil\cr>\cr\sim\cr}}}
{\vcenter{\offinterlineskip\halign{\hfil$\scriptstyle##$\hfil\cr>\cr\sim\cr}}}
{\vcenter{\offinterlineskip\halign{\hfil$\scriptscriptstyle##$\hfil\cr>\cr\sim\cr}}}}}
\def\Co{\mbox{\rm Co}}

\def\Gr{\mbox{\rm Gr}}
\def\Pm{\mbox{\rm Pr}_M}
\def\Rm{\mbox{\rm Re}_M}
\def\Rey{\mbox{\rm Re}}

\def\Co{\mbox{\rm Co}}

\def\cs{c_{\rm s}}
\def\qp{q_{\rm p}}
\def\betastar{\beta_{\star}}
\def\betap{\beta_{\rm p}}

\def\kf{k_{\rm f}}
\def\epsf{\epsilon_{\rm f}}

\def\urms{u_{\rm rms}}

\def\qpz{q_{\rm p0}}
\def\qp{q_{\rm p}}
\def\betap{\beta_{\rm p}}

\def\nut{\nu_{\rm t}}
\def\etat{\eta_{\rm t}}
\def\etatz{\eta_{\rm t0}}

\def\Beq{B_{\rm eq}}
\def\Peff{{\cal P}_{\rm eff}}

\def\half{{\textstyle{1\over2}}}

\def\onethird{{\textstyle{1\over3}}}

\newcommand{\s}{\,{\rm s}}

\newcommand{\etal}{et al.}
\newcommand{\yapj}[3]{ #1, {ApJ,} {#2}, #3}

\newcommand{\yapjl}[3]{ #1, {ApJ,} {#2}, #3}

\newcommand{\yan}[3]{ #1, {Astron.\ Nachr.,} {#2}, #3}

\newcommand{\yana}[3]{ #1, {A\&A,} {#2}, #3}

\newcommand{\yanar}[3]{ #1, {A\&A Rev.,} {#2}, #3}

\newcommand{\ygafd}[3]{ #1, {Geophys.\ Astrophys.\ Fluid Dyn.,} {#2}, #3}

\newcommand{\yjfm}[3]{ #1, {J.\ Fluid Mech.,} {#2}, #3}

\newcommand{\ypf}[3]{ #1, {Phys.\ Fluids,} {#2}, #3}
\newcommand{\ypfb}[3]{ #1, {Phys.\ Fluids B,} {#2}, #3}

\newcommand{\ysovl}[3]{ #1, {Sov.\ Astron.\ Lett.,} {#2}, #3}
\newcommand{\yjetp}[3]{ #1, {Sov.\ Phys.\ JETP,} {#2}, #3}

\newcommand{\yprl}[3]{ #1, {Phys.\ Rev.\ Lett.,} {#2}, #3}

\newcommand{\ymn}[3]{ #1, {MNRAS,} {#2}, #3}

\newcommand{\ysph}[3]{ #1, {Solar Phys.,} {#2}, #3}

\newcommand{\ypre}[3]{ #1, {Phys.\ Rev.\ E,} {#2}, #3}

\newcommand{\pjour}[3]{ #1, {#2}, in press, arXiv:{#3}}
\newcommand{\yjour}[4]{ #1, {#2}, {#3}, #4}

\newcommand{\ybook}[3]{ #1, {#2} (#3)}

\titlerunning{Competing effects in concentrating magnetic flux}
\authorrunning{I. R. Losada \etal}

\title{Competition of rotation and stratification in flux concentrations}
\author{I. R. Losada\inst{1,2,3,4}, A. Brandenburg\inst{1,2},
N. Kleeorin\inst{5,1,6}, I. Rogachevskii\inst{5,1,6}
}
\institute{
Nordita, KTH Royal Institute of Technology and Stockholm University,
Roslagstullsbacken 23, 10691 Stockholm, Sweden
\and
Department of Astronomy, AlbaNova University Center,
Stockholm University, 10691 Stockholm, Sweden
\and
Department of Astrophysics, Universidad de La Laguna,
38206 La Laguna (Tenerife), Spain
\and
Instituto de Astrof\'isica de Canarias, C/ V\'ia L\'actea, s/n, La Laguna,
Tenerife, Spain
\and
Department of Mechanical Engineering, Ben-Gurion University of the Negev,
POB 653, Beer-Sheva 84105, Israel
\and
Department of Radio Physics, N.~I.~Lobachevsky State University of
Nizhny Novgorod, Russia
}
\date{\today,~ $ $Revision: 1.142 $ $}
\begin{document}

\abstract{
In a strongly stratified turbulent layer, a uniform horizontal magnetic
\blue{
field can become unstable to spontaneously form local flux concentrations
}
due to a negative contribution of turbulence to the large-scale (mean-field)
magnetic pressure.
This mechanism, which is called the
negative effective magnetic pressure instability (NEMPI),
is of interest in connection with dynamo scenarios in which
most of the magnetic field resides in the bulk of the convection zone, and
not at the bottom, as is often assumed.
Recent work using the mean-field hydromagnetic equations has shown that
NEMPI becomes suppressed at rather low rotation rates with
Coriolis numbers as low as 0.1.
}{
Here we extend these earlier investigations by studying the effects of
rotation both on the development of NEMPI and on the
effective magnetic pressure.
\blue{
We also quantify the kinetic helicity resulting from
direct numerical simulations (DNS)
with Coriolis numbers and strengths of stratification comparable to
values near the solar surface,
}
and compare with earlier work at smaller scale-separation ratios.
\blue{
Further, we estimate the expected observable signals of magnetic helicity
at the solar surface.
}
}{
To calculate the rotational effect on the effective magnetic pressure
we consider both DNS and analytical studies using the $\tau$ approach.
To study the effects of rotation on the development of NEMPI we use both
DNS and mean-field calculations of the three-dimensional hydromagnetic
equations in a Cartesian domain.
}{
We find that the growth rates of NEMPI from
earlier mean-field calculations are well reproduced with DNS,
provided the Coriolis number is below about 0.06.
\blue{
In that case, kinetic and magnetic helicities are found to be weak and
the rotational effect on the effective magnetic pressure
is negligible as long as the production of flux
concentrations is not inhibited by rotation.
For faster rotation, dynamo action becomes possible.
However, there is an intermediate range of rotation rates
where dynamo action on its own is not yet possible,
but the rotational suppression of NEMPI is being alleviated.
}
}{
Production of magnetic flux concentrations through the suppression of
turbulent pressure appears to be possible only in the upper-most layers
of the Sun, where the convective turnover time is less than 2 hours.
}


\maketitle

\section{Introduction}

In the Sun, magnetic fields are produced by a large-scale dynamo
\citep[see, e.g.,][]{Mof78,Par79,KR80,ZRS83,O03,BS05}.
Although many details of this process remain subject to debate, it seems
relatively clear that rotation enhances the
efficiency of the dynamo if the Coriolis parameter is not very large.
In the absence of rotation and shear, only small-scale magnetic fields are
generated by what is often referred to as small-scale dynamo action
\citep[see, e.g.,][]{ZRS90,BS05}.
Rotation leads to an $\alpha$ effect \citep{SKR66}
if there is also stratification in density or turbulent intensity.
The $\alpha$ effect can produce mean magnetic field and net magnetic flux.

Stratification leads to yet another effect which does not
produce magnetic flux, but merely concentrates it locally by what is now
referred to as negative effective magnetic pressure instability (NEMPI).
Direct numerical simulations (DNS) of \cite{BKKMR11} have shown in
surprising detail many aspects of NEMPI that were previously seen in
mean-field simulations (MFS) of \cite{BKR10} and that have been
anticipated based on analytical studies for some time
\citep{KRR89,KRR90,KMR93,KMR96,KR94,RK07}.

The main physics of this effect is connected with the suppression of
turbulent pressure by a weak
mean magnetic field that is less than the equipartition field.
At large Reynolds numbers, the resulting reduction of the turbulent
pressure is larger than the added magnetic pressure from the mean
magnetic field itself,
so that the {\em effective} magnetic pressure
that accounts for turbulent and non-turbulent contributions,
becomes negative.
In a strongly stratified layer, i.e., a layer in which the density
varies much more rapidly with height than the magnetic field, this leads
to an instability that is analogous to Parker's magnetic buoyancy
instability, except that
there the magnetic field varies more rapidly with height than the density.
Because the effective magnetic pressure is negative, magnetic structures
are negatively buoyant and sink, which has been seen
in DNS of \cite{BKKMR11}.

One of the main successes of recent comparative
work between DNS and MFS is the demonstration of
a high degree of predictive power of MFS. The
examples include details regarding the shape and
evolution of structures, the dependence of their
depth on the magnetic field strength, and the
dependence of the growth rate on the scale
separation ratio.
Recent MFS of \cite{Losada}
(hereafter LBKMR) have shown that in the presence
of even just weak rotation, the growth rate of
NEMPI is significantly reduced. Expressed in
terms of the Coriolis number,
$\Co=2\Omega/\urms\kf$, where $\Omega$ is the
angular velocity, $\urms$ is the rms velocity of
the turbulence, and $\kf$ is the wavenumber of
the energy-carrying eddies, the critical value of
$\Co$ was predicted to be as low as 0.03.
Although this value does not preclude the
operation of NEMPI in the upper parts of the Sun,
where $\Co$ is indeed small (about $10^{-4}$ at the surface),
it does seem surprisingly low, which raises questions
regarding the accuracy of MFS in this case. The
purpose of the present paper is therefore to
compare MFS of LBKMR with DNS of the same setup.
It turns out that, while we do confirm the basic prediction
of LBKMR, we also resolve an earlier noticed discrepancy
in the growth rates between DNS and MFS in the absence
of rotation \citep[see the appendix of][]{KBKMR12c}.
Indeed, in the particular case of a magnetic Reynolds number
of 18 and a scale separation ratio of 30, the formation of
structures is unusually strong and the averaged stratification
\blue{
changes significantly to affect the determination of the effective
}
magnetic pressure.
However, by restricting the analysis to early times, we obtain
coefficients that are not only in better agreement with an earlier
formula of \cite{BKKR12} with a smaller scale separation ratio,
but that also give MFS results that agree better with our new DNS.

The DNS are used primarily to compute the growth rates
and magnetic field structures during the saturated state
without invoking the mean-field concept at all.
By contrast, the $\tau$ approach \citep{O70,PFL76,KRR90,RK04}
is used to determine the
dependence of mean-field coefficients on the rotation rate.
This can also be done with DNS \citep{KBKMR12c}.
Here we apply those calculations to the case with rotation.

We recall that we adopt here an isothermal stratification and an
isothermal equation of state.
This is done because the effect that we are interested in exists even in
this simplest case where temperature and pressure scale height are constant.
Non-isothermal setups have been studied at the mean-field level both with
\citep{Kapy12,KIAU13} and without
\citep{BKR10} entropy evolution included.
In a stably stratified layer, entropy evolution leads to an additional
restoring force and hence to
internal gravity waves (Brunt-V\"ais\"al\"a oscillations)
that stabilize NEMPI \citep{Kapy12}.
Thus, by using both isothermal stratification and an isothermal equation
of state, we recover a situation that is similar to an adiabatic layer,
except that then the temperature and hence the pressure scale height
decrease with height.

The system we are thus dealing with is governed by the combined action
of rotation and stratification.
In principle, such systems have been studied many times before, for
example to determine the $\alpha$ effect in mean-field dynamo theory
\citep{KR80,BS05}.  The difference to earlier work is the large scale
separation ratio, where the domain is up to 30 times larger than the
scale of the energy-carrying eddies.
As mentioned in the beginning, stratification and rotation lead to
kinetic helicity and an $\alpha$ effect.
We therefore also quantify here the amount of kinetic helicity produced
and whether this leads to observable effects in the resulting magnetic
structures.
We use here the opportunity to explore the feasibility of determining
the magnetic helicity spectrum from measurements of the magnetic correlation
tensor along a longitudinal strip.

We begin by discussing first the basic equations to determine the
effective magnetic pressure from DNS and the $\tau$ approach (Section~2),
compare growth rates for MFS and DNS (Section~4), and turn then to the
measurement of kinetic and magnetic helicity from surface measurements
(Section~5), before concluding in Section~6.

\section{The model}

We consider DNS of an isothermally stratified layer \citep{BKKMR11,KBKMR12c}
and solve the equations for the velocity $\UU$,
the magnetic vector potential $\AAA$, and the density $\rho$,
in the presence of rotation $\Omega$,
\begin{eqnarray}
{\DD\UU\over\DD t}&=&-2\OO\times\UU
-\cs^2\nab\ln\rho+{1\over\rho}\JJ\times\BB+\ff+\grav+\FF_\nu,\\
{\partial\AAA\over\partial t}&=&\UU\times\BB+\eta\nabla^2\AAA,\\
{\partial\rho\over\partial t}&=&-\nab\cdot\rho\UU,
\end{eqnarray}
where $\DD/\DD t=\partial/\partial t+\UU\cdot\nab$ is the advective
derivative, $\nu$ is the kinematic viscosity, $\eta$ is the
magnetic diffusivity due to Spitzer conductivity of the plasma,
$\BB=\BB_0+\nab\times\AAA$ is the magnetic field,
$\BB_0=(0,B_0,0)$ is the imposed uniform field,
$\JJ=\nab\times\BB/\mu_0$ is the current density,
$\mu_0$ is the vacuum permeability,
$\FF_\nu=\nab\cdot(2\nu\rho\SSSS)$ is the viscous force, ${\sf S}_{ij}
=\half(\partial_j U_i+\partial_i U_j)-\onethird\delta_{ij}\nab\cdot\UU$
is the traceless rate-of-strain tensor.
The angular velocity vector $\OO$ is quantified by its scalar amplitude
$\Omega$ and colatitude $\theta$, such that
$\OO=\Omega\left(-\sin\theta, 0, \cos\theta\right)$.
As in LBKMR, $z$ corresponds to radius,
$x$ to colatitude, and $y$ to azimuth.
The forcing function $\ff$ consists of random, white-in-time,
plane, non-polarized waves with a certain average wavenumber $\kf$.
The turbulent rms velocity is approximately
independent of $z$ with $\urms=\bra{\uu^2}^{1/2}\approx0.1\,\cs$.
The gravitational acceleration $\grav=(0,0,-g)$ is chosen such that
$k_1 H_\rho=1$, so the density contrast between
bottom and top is $\exp(2\pi)\approx535$
in a domain $-\pi\leq k_1 z\leq\pi$.
Here, $H_\rho=\cs^2/g$ is the density scale height
and $k_1=2\pi/L$ is the smallest wavenumber that fits into
the cubic domain of size $L^3$.
In most of our calculations, structures develop whose
horizontal wavenumber $k_x$ is close to $k_1$.
We adopt Cartesian coordinates $(x,y,z)$,
with periodic boundary conditions in
the $x$- and $y$-directions and stress-free, perfectly conducting
boundaries at the top and bottom ($z=\pm L_z/2$).
In all cases, we use a scale separation ratio $\kf/k_1$ of 30,
a fluid Reynolds number $\Rey\equiv\urms/\nu\kf$ of 36,
and a magnetic Prandtl number $\Pm=\nu/\eta$ of 0.5.
The magnetic Reynolds number is therefore $\Rm=\Pm\Rey=18$.
The value of $B_0$ is specified in units of the volume-averaged
value $\Beqz=\sqrt{\mu_0\rho_0} \, \urms$,
where $\rho_0=\bra{\rho}$ is the volume-averaged density,
which is constant in time.
As in earlier work, we also define the local equipartition
field strength $\Beq(z)=\sqrt{\mu_0\rho} \, \urms$.
In our units, $k_1=\cs=\mu_0=\rho_0=1$.
In addition to visualizations of the actual magnetic field,
we also monitor $\meanB_y$, which is an
average over $y$ and a certain time interval $\Delta t$.
Time is sometimes specified in terms of turbulent-diffusive times
$t\,\etatz k_1^2$, where $\etatz=\urms/3\kf$ is the estimated
turbulent diffusivity.

The simulations are performed with the {\sc Pencil Code}
\url{http://pencil-code.googlecode.com} which uses sixth-order explicit
finite differences in space and a third-order accurate time stepping method.
We use a numerical resolution of $256^3$ mesh points.

We compare with and extend earlier MFS of LBKMR, where
we solve the evolution equations for mean velocity $\meanUU$,
mean density $\meanrho$, and mean vector potential $\meanAA$, in the form
\EQA
\label{dUmean}
{\partial\meanUU\over\partial t}&=&-\meanUU\cdot\nab\meanUU-2\OO\times\meanUU
-\cs^2\nab\ln\meanrho+\grav+\meanFFFF_{\rm MK},
\\
{\partial\meanAA\over\partial t}&=&\meanUU\times\meanBB-(\etat+\eta)\meanJJ, \\
{\partial\meanrho\over\partial t}&=&-\meanUU\cdot\nab\meanrho
-\meanrho\nab\cdot\meanUU,
\ENA
where $\meanFFFF_{\rm MK}=\meanFFFF_{\rm M}+\meanFFFF_{\rm K}$, with
\EQ
\meanrho \, \meanFFFF_{\rm M} = -\half\nab[(1-q_{\rm p})\meanBB^2]
\label{efforce}
\EN
being the mean-field magnetic pressure force, and
\EQ
\meanFFFF_{\rm K}=(\nut+\nu)\left(\nabla^2\meanUU+\onethird\nab\nab\cdot\meanUU
+2\meanSSSS\nab\ln\meanrho\right)
\EN
is the total (turbulent plus microscopic) viscous force.
Here, $\meanSSS_{ij}=\half(\meanU_{i,j}+\meanU_{j,i})
-\onethird\delta_{ij}\nab\cdot\meanUU$
is the traceless rate-of-strain tensor of the mean flow
and $\qp$ is approximated by \citep{KBKR12}
\EQ
\qp(\beta)={\betastar^2\over\betap^2+\beta^2},
\label{qp}
\EN
which is only a function of the ratio $\beta\equiv|\meanBB|/\Beq(z)$.
Here, $\betastar$ and $\betap$ are coefficients that have been determined
from previous numerical simulations in the absence of rotation \citep{BKKR12}.
In \Eq{efforce} we have taken into account that the mean magnetic field
is independent of $y$, so the mean magnetic tension vanishes.

The strength of gravitational stratification is characterized by
the nondimensional parameter $\Gr=g/\cs^2\kf \equiv (H_\rho \kf)^{-1}$
\citep{BRK12}.
Another important nondimensional parameter is the Coriolis number,
$\Co=2\Omega/\urms\kf$.
Alternatively, we normalize the growth rate of the instability by a quantity
\EQ
\lambda_{\ast0}\equiv\betastar\urms/H_\rho,
\EN
which is motivated by the analytic results of LBKMR and the finding that
NEMPI is suppressed when $2\Omega\ga\lambda_{\ast0}$.

\section{Effective magnetic pressure}

In this section we study the effect of rotation
on the function $\qp(\beta)$.
We consider first the results of DNS and turn then to
an analytical treatment.

\subsection{Numerical results}
\label{NumericalResults}

\begin{figure}[t!]\begin{center}
\includegraphics[width=\columnwidth]{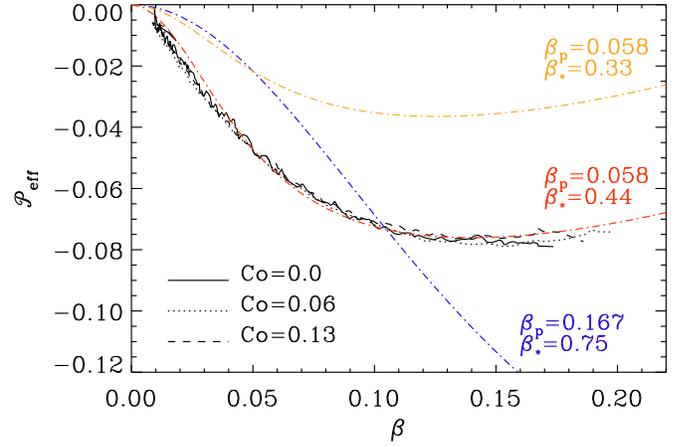}
\end{center}\caption[]{
Normalized effective magnetic pressure, $\Peff(\beta)$, for
three values of $\Co$, compared with \Eq{qp} for different
combinations of $\betastar$ and $\betap$, as discussed in the text.
}\label{ppresseff}\end{figure}

In the MFS of LBKMR we assumed that $\Peff(\beta)$ does not change
significantly with $\Co$ in the range considered.
With DNS we can compute $\Peff(\beta)$ by calculating the combined
Reynolds and Maxwell stress for a run with and one without an imposed
magnetic field.
This allows us to compute $\qp(\beta)$ using Eq.~(17) of \cite{BKKR12}:
\EQ
\qp=-2 \left. \left[\meanrho\,(\overline{u_x^2}-\overline{u_{0x}^2})
+\half\overline{\bb^2}-\overline{b_x^2} \right] \right/ \meanBB^2,
\EN
where the subscripts 0 indicate values obtained from a reference
run with $B_0=0$. This expression does not take into account
small-scale dynamo action which can produce finite background
magnetic fluctuations $\bb_0$.
The effective magnetic pressure is then determined using the equation
$\Peff(\beta)=\half[1-\qp(\beta)]\beta^2$.
The result is plotted in \Fig{ppresseff} for three values of $\Co$
during an early time interval when structure formation is still weak
and the background stratification remains unchanged so that the
result is not yet affected.
\blue{
We note that even in the $\Co=0.13$ case, in which the instability
is no longer so prominent,
we have to restrict ourselves to early times,
since the negative effective magnetic pressure affects the background
stratification and hence the pressure changes at later times.
The resulting profiles of $\Peff(\beta)$ are virtually
}
the same for all three values of $\Co$.
We also compare with \Eq{qp} for different combinations of
$\betastar$ and $\betap$.
It turns out that the curves for different values of $\Co$ are best
reproduced for $\betastar=0.44$ and $\betap=0.058$.

\subsection{Theoretical predictions}
\label{theory}

Let us now compare with theoretical predictions for $\qp(\beta)$.
We take into account the feedback of the magnetic field on the
turbulent fluid flow. We use a mean-field
approach whereby velocity, pressure and magnetic
field are separated into mean and fluctuating
parts. We also assume vanishing mean motion. The
strategy of our analytic derivation is to
determine the $\Omega$ dependencies of the second
moments for the velocity $\overline{u_{i}(t,{\bm
x}) \, u_{j}(t,{\bm x})}$, the magnetic field
$\overline{b_i(t,{\bm x}) \, b_{j}(t,{\bm x})}$,
and the cross-helicity tensor
$\overline{b_i(t,{\bm x}) \, u_{j}(t,{\bm x})}$,
where ${\bm b}$ are fluctuations of magnetic
field produced by tangling of the large-scale
field. To this end we use the equations for
fluctuations of velocity and magnetic field in
rotating turbulence, which are obtained by
subtracting equations for the mean fields from
the corresponding equations for the actual (mean
plus fluctuating) fields.

\subsubsection{Governing equations}

The equations for the fluctuations of velocity
and magnetic fields are given by
\begin{eqnarray}
{\partial {\bm u}({\bm x},t) \over \partial t} \!&=&\!
{1 \over \meanrho} \left( \meanBB \cdot \bec{\nabla}{\bm b}
+ {\bm b} \cdot \bec{\nabla} \meanBB -\bec{\nabla} p \right)
+ 2 {\bm u} \times {\bm \Omega} + \hatNNN^u,
\nonumber\\
 \label{B1} \\
{\partial {\bm b}({\bm x},t) \over \partial t}
\!&=&\! \meanBB \cdot \bec{\nabla}{\bm u}
- {\bm u} \cdot \bec{\nabla} \meanBB + \hatNNN^b,
 \label{B2}
\end{eqnarray}
where Eq.~(\ref{B1}) is written in a reference
frame rotating with constant angular velocity
${\bm \Omega}$, $\, p = p' + (\meanBB \cdot {\bm
b})$ are the fluctuations of total pressure, $p'$
are the fluctuations of fluid pressure, $\meanBB$
is the mean magnetic field, and $\meanrho$ is the mean fluid
density. For simplicity we neglect effects of
compressibility. The terms $\hatNNN^u$
and $\hatNNN^b$, which include
nonlinear and molecular viscous and dissipative
terms, are given by
\begin{eqnarray}
\hatNNN^u
\!&=&\! \overline{{\bm u} \cdot \bec{\nabla} {\bm u}} - {\bm u} \cdot
\bec{\nabla} {\bm u} +{1\over\meanrho} \left({\bm j}
\times {\bm b} - \overline{{\bm j} \times {\bm
b}}\right)
+ {\bm f}_{\nu}({\bm u}),
\\
\hatNNN^b
\!&=&\! \bec{\nabla} \times
\left({\bm u} \times {\bm b} - \overline{{\bm u}
\times {\bm b}} - \eta \bec{\nabla} \times {\bm
b} \right),
\end{eqnarray}
where $\meanrho {\bm f}_{\nu}({\bm u})$ is the
molecular viscous force and ${\bm j}=\nab\times{\bm b}/\mu_0$
is the fluctuating current density.
To eliminate the pressure term from the equation of
motion~(\ref{B1}) we calculate $\bec{\nabla} {\bm
\times} (\bec{\nabla} {\bm \times} {\bm u})$.
Then we rewrite the obtained equation and \Eq{B2}
in Fourier space.

\subsubsection{Two-scale approach}

We apply the two-scale approach and express two-point
correlation functions in the following form
\begin{eqnarray}
\overline{ u_i ({\bm x}) u_j ({\bm  y}) }\!&=&\!\int
\,d{\bm k}_1 \, d{\bm  k}_2 \, \overline{
u_i({\bm  k}_1) u_j ({\bm k}_2) } \exp \{i({\bm
k}_1 {\bm \cdot} {\bm x}
+ {\bm  k}_2 {\bm \cdot} {\bm y}) \}
\nonumber \\
&=&\!\int \,d {\bm  k} \,d {\bm  K} \, f_{ij}({\bm k,
K}) \exp (i {\bm k} {\bm \cdot} {\bm r}+ i {\bm
K} {\bm \cdot} {\bm R})
\nonumber \\
&=&\!\int  \,d {\bm  k} \, f_{ij}({\bm k,
R}) \exp (i {\bm k} {\bm \cdot}{\bm r})
\end{eqnarray}
\citep[see, e.g.,][]{RS75}. Here and elsewhere, we drop the common
argument $t$ in the correlation functions,
$f_{ij}({\bm k, R}) = \hat L(u_i; u_j)$, where
\begin{eqnarray*}
\hat L(a; c) = \int \overline{ a({\bm k} + {\bm
K} / 2) c(-{\bm k} + {\bm  K} / 2) } \exp{(i {\bm
K} {\bm \cdot} {\bm R}) } \,d {\bm  K},
\end{eqnarray*}
with the new variables ${\bm R} = ({\bm
x} +  {\bm y}) / 2,$ $\, {\bm r} = {\bm x} - {\bm
y}$, $\, {\bm K} = {\bm k}_1 + {\bm k}_2$, ${\bm
k} = ({\bm k}_1 - {\bm k}_2) / 2$. The variables
${\bm R}$ and ${\bm K}$ correspond to the large
scales, while ${\bm r}$ and ${\bm k}$ correspond
to the  small scales. This implies that we have
assumed that there exists a separation of scales,
i.e., the turbulent forcing scale $\ell_{\rm f}$ is
much smaller than the characteristic scale $L_B$
of inhomogeneity of the mean magnetic field.

\subsubsection{Equations for the second moments}

We derive equations for the following correlation
functions: $f_{ij}({\bm k, R}) = \hat L(u_i;
u_j)$, $\, h_{ij}({\bm k, R}) = \meanrho^{-1} \,
\hat L(b_i; b_j)$ and $g_{ij}({\bm k, R}) = \hat
L(b_i; u_j)$. The equations for these correlation
functions are given by
\begin{eqnarray}
&& {\partial f_{ij}({\bm k}) \over \partial t} =
i({\bm k} {\bm \cdot} \overline{\bm B}) \Phi_{ij}
+ L_{ijmn}^{\Omega} f_{mn} + I^f_{ij}
+ \hatNN_{ij}^f  ,
\label{B6} \\
&& {\partial h_{ij}({\bm k}) \over \partial t} =
- i({\bm k}{\bm \cdot} \overline{\bm B})
\Phi_{ij} + I^h_{ij}
+ \hatNN_{ij}^h  ,
\label{B7} \\
&& {\partial g_{ij}({\bm k}) \over \partial t} =
i({\bm k} {\bm \cdot} \overline{\bm B})
[f_{ij}({\bm k}) - h_{ij}({\bm k}) -
h_{ij}^{(H)}]
 \nonumber\\
&& \quad + D_{jm}^{\Omega}({\bm k}_2)g_{im}({\bm
k}) + I^g_{ij}
+ \hatNN_{ij}^g  ,
 \label{B8}
\end{eqnarray}
where
\begin{eqnarray*}
\Phi_{ij}({\bm k}) &=& \meanrho^{-1}\, [g_{ij}({\bm
k}) - g_{ji}(-{\bm k})],
\\
D_{ij}^{\Omega}({\bm k}) &=& 2 \varepsilon_{ijm} \Omega_n k_{mn},
\\
L_{ijmn}^{\Omega} &=& D_{im}^{\Omega}({\bm k}_1) \, \delta_{jn} +
D_{jn}^{\Omega}({\bm k}_2) \, \delta_{im} .
\end{eqnarray*}
Hereafter we have omitted the ${\bm R}$-argument in the
correlation functions and neglected terms $\sim
O(\nabla_{\bm R}^2)$,
and $\varepsilon_{ijn}$ is the fully antisymmetric
Levi-Civita tensor.
In Eqs.~(\ref{B6})--(\ref{B8}),
the terms $\hatNNN^f$,
$\hatNNN^h$ and $\hatNNN^g$
are determined by the third moments appearing
due to the nonlinear
terms, the source terms $I_{ij}^f$ , $\,
I_{ij}^h$ and $I_{ij}^g$ which contain the
large-scale spatial derivatives of the mean
magnetic and velocity fields, are given by
Eqs.~(A3)--(A6) in \cite{RK04}. These terms
determine turbulent magnetic diffusion and
effects of nonuniform mean velocity on the mean
electromotive force.

For the derivation of Eqs.~(\ref{B6})--(\ref{B8})
we use an approach that is similar to that
applied in \cite{RK04}.
We take into account that in Eq.~(\ref{B8}) the terms
with tensors that are symmetric in $i$ and $j$ do not
contribute to the mean electromotive force
because ${\cal E}_{m} = \varepsilon_{mji} \,
g_{ij}$. We split all tensors into nonhelical,
$h_{ij},$ and helical, $h_{ij}^{(H)},$ parts. The
helical part of the tensor of magnetic
fluctuations $h_{ij}^{(H)}$ depends on the
magnetic helicity, and the equation for
$h_{ij}^{(H)}$ follows from magnetic helicity
conservation arguments \citep[see, e.g.,][and
references therein]{KR99,BS05}.

\subsubsection{$\tau$-approach}
\label{Tau}

The second-moment
equations~(\ref{B6})--(\ref{B8}) include the
first-order spatial differential operators
applied to the third-order moments $M^{\rm(III)}$.
To close the system, we express the set of
the third-order terms
$\hatNNN^M \equiv \hatNNN M^{\rm(III)}$
through the lower moments $M^{\rm(II)}$. We use
the spectral $\tau$ approximation which
postulates that the deviations of the
third-moment terms, $\hatNNN M^{\rm(III)}({\bm k})$,
from the contributions to these terms afforded by the background
turbulence, $\hatNNN M^{\rm(III,0)}({\bm
k})$, are expressed through similar
deviations of the second moments:
\begin{eqnarray}
&& \hatNNN M^{\rm(III)}({\bm k}) - \hatNNN
M^{\rm(III,0)}({\bm k}) =- {M^{\rm(II)}({\bm
k}) - M^{\rm(II,0)}({\bm k}) \over \tau(k)}
\nonumber\\
\label{T10}
\end{eqnarray}
\citep{O70,PFL76,KRR90,RK04}, where $\tau(k)$ is
the scale-dependent relaxation time, which can be
identified with the correlation time of the
turbulent velocity field for large Reynolds numbers.
The quantities with the
superscript $(0)$ correspond to the background
turbulence (see below). We apply the spectral
$\tau$ approximation only for the nonhelical part
$h_{ij}$ of the tensor of magnetic fluctuations.
A justification for the $\tau$ approximation in
different situations has been offered through
numerical simulations and analytical studies
\cite[see, e.g.,][]{BKM04,BS05b,BS07,RKB11}.

\subsubsection{Solution of equations for the second moments}

We solve Eqs.~(\ref{B6})--(\ref{B8})
neglecting the sources $I^f_{ij}, I^h_{ij},
I^g_{ij}$ with the large-scale spatial
derivatives. The terms with the large-scale
spatial derivatives which determine the turbulent
magnetic diffusion, can be taken into account by
perturbations. We subtract from
Eqs.~(\ref{B6})--(\ref{B8}) the corresponding
equations written for the background turbulence,
use the spectral $\tau$ approximation. We assume
that the characteristic time of variation of the
second moments is substantially larger than the
correlation time $\tau(k)$ for all turbulence
scales. This allows us to get a stationary
solution for the equations for the second-order
moments, $M^{\rm(II)}$. Thus, we arrive to the
following steady-state solution of
Eqs.~(\ref{B6})--(\ref{B8}):
\begin{eqnarray}
&& f_{ij}({\bm k}) = L_{ijmn}^{-1}
\left[f_{mn}^{(0)}({\bm k}) + i \tau ({\bm k}
{\bm \cdot} \overline{\bm B}) \Phi_{mn}({\bm
k})\right] ,
 \label{B17}\\
&& h_{ij}({\bm k}) = - i \tau({\bm k} {\bm \cdot}
\overline{\bm B}) \Phi_{ij}({\bm k}),
 \label{B18}\\
&& g_{ij}({\bm k}) = i \tau ({\bm k} {\bm \cdot}
\overline{\bm B}) D_{im}^{-1} \left[f_{mj}({\bm
k}) - h_{mj}({\bm k})\right].
 \label{B19}
\end{eqnarray}
We have assumed that there is no small-scale dynamo in
the background turbulence. Here the operator
$D_{ij}^{-1}$ is the inverse of the operator
$\delta_{ij} - \tau D_{ij}^{\Omega}$ \citep{RKR03} and the
operator $L_{ijmn}^{-1}$ is the inverse of the
operator $\delta_{im} \delta_{jn} - \tau \,
L_{ijmn}^{\Omega}$ \citep{EKR05}. These operators are given by
\begin{eqnarray}
&& D_{ij}^{-1} = \chi(\psi) \, (\delta_{ij} +
\psi \, \varepsilon_{ijm} \, \hat k_m + \psi^2 \,
k_{ij})
\nonumber\\
&& \quad = \delta_{ij} + \psi \, \varepsilon_{ijm} \, \hat k_m - \psi^2 \,
P_{ij} + O(\psi^3),
\label{B12}\\
&& L_{ijmn}^{-1}({\bm \Omega}) = \half [B_1
\, \delta_{im} \delta_{jn} + B_2 \, k_{ijmn} +
B_3 \, (\varepsilon_{imp} \delta_{jn}
\nonumber\\
&& \quad + \varepsilon_{jnp} \delta_{im}) \hat k_p +
B_4 \, (\delta_{im} k_{jn} + \delta_{jn} k_{im})
\nonumber\\
&& \quad + B_5 \, \varepsilon_{ipm} \varepsilon_{jqn}
k_{pq} + B_6 \, (\varepsilon_{imp} k_{jpn} +
\varepsilon_{jnp} k_{ipm}) ]
\nonumber\\
&& \quad = \delta_{im} \delta_{jn} + \psi \, (\varepsilon_{imp} \delta_{jn}
+ \varepsilon_{jnp} \delta_{im}) \hat k_p - \psi^2
(\delta_{im} P_{jn}
\nonumber\\
&& \quad + \delta_{jn} P_{im}
- 2 \varepsilon_{imp} \varepsilon_{jqn} k_{pq}) + O(\psi^3),
\label{B14}
\end{eqnarray}
where $\hat k_i = k_i / k$, $\, \chi(\psi) = 1 / (1
+ \psi^2) $, $\, \psi = 2 \tau(k) \, ({\bm k}
\cdot {\bm \Omega}) / k $, $\, B_1 = 1 + \chi(2
\psi) ,$ $\, B_2 = B_1 + 2 - 4 \chi(\psi) ,$ $\,
B_3 = 2 \psi \, \chi(2 \psi) ,$ $\, B_4 = 2
\chi(\psi) - B_1 ,$ $\, B_5 = 2 - B_1 $ and $B_6
= 2 \psi \, [\chi(\psi) - \chi(2 \psi)]$,
$P_{ij}(k) =\delta_{ij} - k_i k_j / k^2$,
$\delta_{ij}$ is the Kronecker tensor.

We use the following model for the homogeneous
and isotropic background turbulence:
$f_{ij}^{(0)}({\bm k})= \langle {\bm u}^2 \rangle
\, P_{ij}(k) \, W(k)$, where $W(k) = E(k) / 8 \pi k^{2}
$, the energy spectrum is $E(k) = (q-1)
k_{0}^{-1} (k / k_{0})^{-q}$, $k_{0} = 1 /
\ell_{\rm f}$ and the length $\, \ell_{\rm f}$ is the
maximum scale of turbulent motions. The turbulent
correlation time is $\tau(k) = C \, \tau_0 \, (k
/ k_{0})^{-\mu}$, where the coefficient
$C=(q-1+\mu)/(q-1)$. This value of the
coefficient $C$ corresponds to the standard form
of the turbulent diffusion coefficient in the
isotropic case, i.e., $\eta_{_{T}} = \langle {\bm u}^2 \rangle
\int \tau(k) \, E(k)\, dk =
\tau_0 \, \langle {\bm u}^2 \rangle /3$. Here the
time $\tau_0 = \ell_{\rm f} / \sqrt{\langle {\bm u}^2
\rangle}$ and $\sqrt{\langle {\bm u}^2 \rangle}$
is the characteristic turbulent velocity in the
scale $\ell_{\rm f}$. For the Kolmogorov's type
background turbulence (i.e., for a turbulence
with a constant energy flux over the spectrum),
the exponent $\mu=q-1$ and the coefficient $C=2$.
In the case of a turbulence with a
scale-independent correlation time, the exponent
$\mu=0$ and the coefficient $C=1$. Motions in the
background turbulence are assumed to be
non-helical.

Equations~(\ref{B17})--(\ref{B14}) yield:
\begin{eqnarray}
&& f_{ij}({\bm k}) = f_{ij}^{(0)}({\bm k}) - h_{ij}({\bm k}),
 \label{B20}\\
&& h_{ij}({\bm k}) = {\Psi \over 1+2\Psi}
\left(1 - \psi^2 {2+ \Psi \over 2(1+2\Psi)}\right)
f_{ij}^{(0)}({\bm k}),
 \label{B21}
\end{eqnarray}
where $\Psi=2 \tau^2 ({\bm k} {\bm \cdot} {\bm c}_A)^2$,
${\bm c}_A=\overline{\bm B} / \sqrt{\meanrho}$,
and we have taken into account that
$L^{-1}_{ijmn} P_{mn}({\bm k}) = P_{ij}({\bm k})$.
After the integration in ${\bm k}$ space
we obtain the magnetic tensor $h_{ij}$ in physical space:
\begin{eqnarray}
&& h_{ij}(\beta) = q_1(\beta) \delta_{ij} + q_2(\beta)
\beta_{ij} ,
\label{B22}
\end{eqnarray}
where $\beta =\overline{B} / B_{\rm eq}$, and
the functions $q_1(\beta)$ and $q_2(\beta)$ are given
in Appendix~A.
We consider the case in which the angular velocity
is perpendicular to the mean magnetic field.
The results can easily be generalized to the case of
the arbitrary angle between the angular velocity and
the mean magnetic field.

The contribution of turbulence to the mean-field magnetic
pressure is given by the function $q_{\rm p}(\beta)=\left[q_1(\beta)
- q_2(\beta)\right] / \beta^2$:
\begin{eqnarray}
&& q_{\rm p}(\beta) = {1 \over 12 \beta^2} \Big[A_1^{(0)}(0)
- A_1^{(0)}(4\beta) - A_2^{(0)}(4\beta)
\nonumber\\
&& \quad - 2(\Omega \tau_0)^2 \Big(A_1^{(2)}(0)
 - 4C_1^{(2)}(0) - 10 A_1^{(2)}(4\beta)
\nonumber\\
&& \quad + 40 C_1^{(2)}(4\beta)
+ {9 \over 2 \pi} \left[\bar A_1(16\beta^2) -4 \bar C_1(16\beta^2)
\right] \Big) \Big],
\label{B25}
\end{eqnarray}
where the functions $A_i^{(j)}(x)$,
$C_i^{(j)}(x)$, $\bar A_i(y)$ and $\bar C_i(y)$,
and their asymptotics are given in Appendix~A.
Following earlier work \citep{BKKR12}, we now define a magnetic
Reynolds number based on the scale $\ell_{\rm f}=2\pi/\kf$,
which is related to the $\Rm$ defined earlier via ${\rm Rm}=2\pi\Rm$.
For $\overline{B} \ll B_{\rm eq} / 4 {\rm Rm}^{1/4}$,
the function $q_{\rm p}(\beta)$ is given by
\begin{eqnarray}
q_{\rm p}(\beta) = {4 \over 5} \ln {\rm Rm} - {8 \over 35} \Co^2,
\label{B26}
\end{eqnarray}
and for $B_{\rm eq} / 4 {\rm Rm}^{1/4} \ll
\overline{B} \ll B_{\rm eq} / 4$ the function $q_{\rm p}(\beta)$
is given by
\begin{eqnarray}
q_p(\beta) &=& {16 \over 25} \, \left(1 + 5|\ln (4 \beta)| + 32
\, \beta^{2} \right) - {8 \over 35} \Co^2,
\label{B27}
\end{eqnarray}
where $\Co = 2 \Omega \tau_0$.
This shows that for the values of $\Co$ of interest ($\Co\le0.06$),
the correction to $\qp$ is negligible (below $10^{-3}$), which is
in agreement with the numerical findings in \Fig{ppresseff}.

\section{Coriolis effects of NEMPI in DNS and MFS}

\subsection{DNS and comparison with MFS}
\label{DNSandMFS}

\begin{figure}[t!]\begin{center}
\includegraphics[width=\columnwidth]{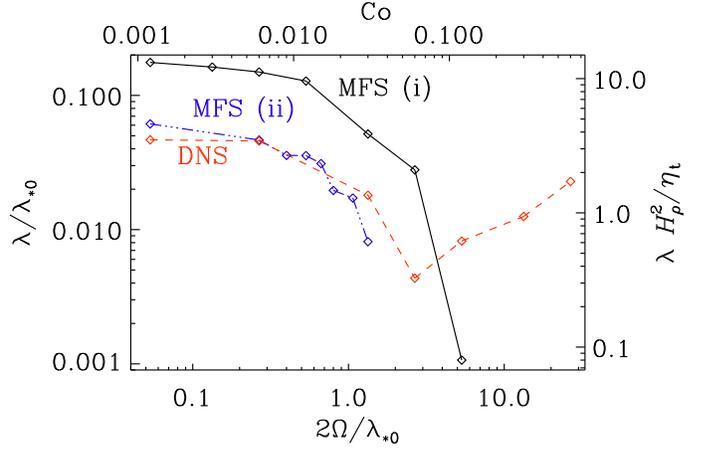}
\end{center}\caption[]{
Dependence of $\lambda/\lambda_{\ast0}$ on $2\Omega/\lambda_{\ast0}$
for DNS (red dashed line), compared with MFS~(i) where $\qpz=20$ and
$\betap=0.167$ (black solid line), as well as MFS~(ii) where $\qpz=32$
and $\betap=0.058$ (blue dash-dotted line).
In this case no growth was found for $\Co\ge0.03$.
In all cases we have $B_0/\Beqz=0.05$.
}\label{presults_Omdep}\end{figure}

\begin{figure*}[t]\begin{center}
\includegraphics[width=\textwidth]{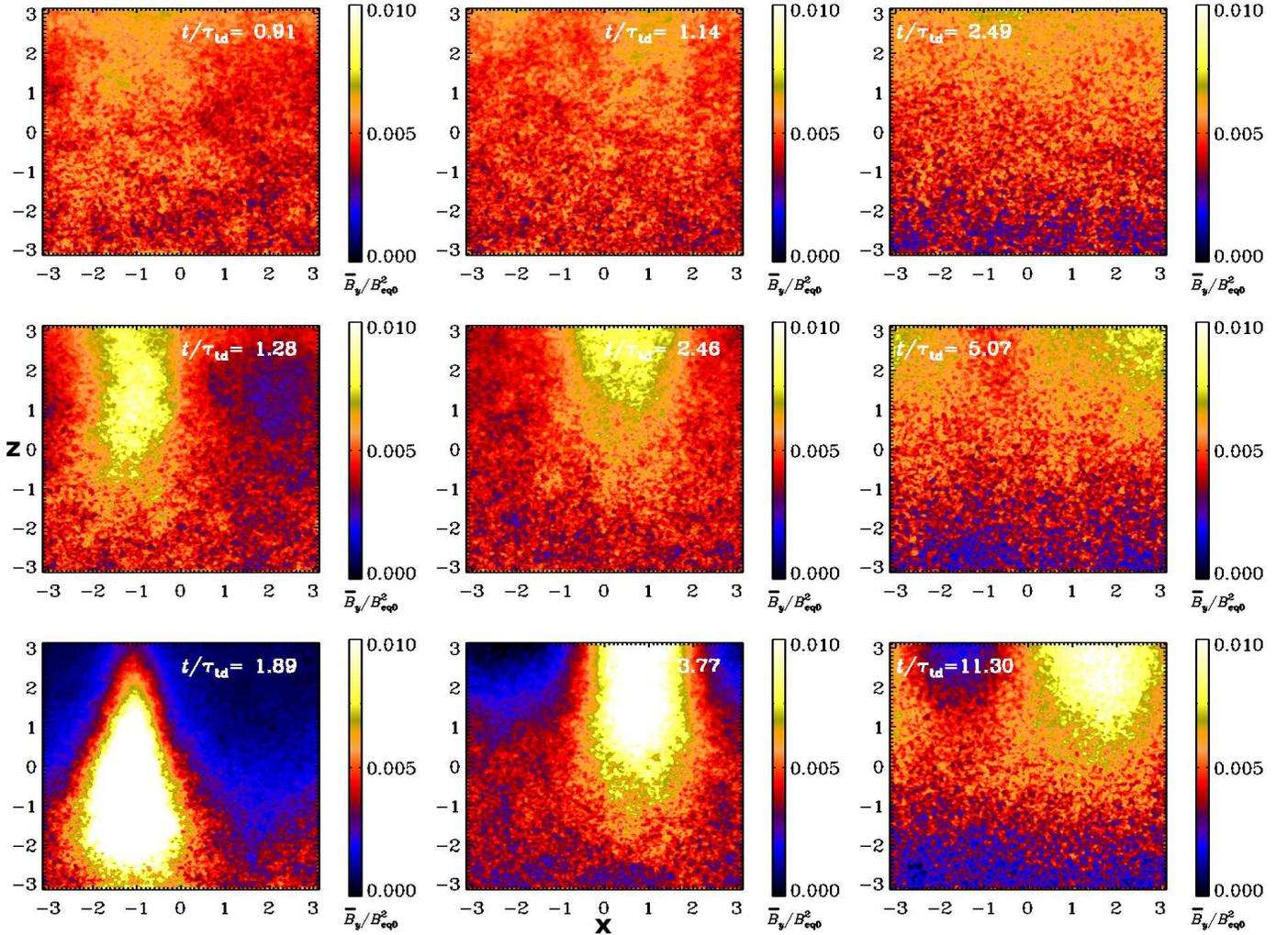}
\end{center}\caption[]{
$yt$-averaged $\meanB_y$ for $\Co=0.006$ (left), 0.03 (middle),
and 0.06 (right) at different times.
}\label{b005_th0}\end{figure*}

We have performed DNS for different values of $\Co$ and calculated
the growth rate $\lambda$; see \Fig{presults_Omdep}.
\blue{
It turns out that $\lambda$ shows a decline with increasing values
of $\Co$ that is similar to that seen in the MFS of LBKMR, who used
$\qpz=20$ and $\betap=0.167$ (corresponding to $\betastar=0.75$).
However, for $\Co=0.13$ and 0.66, some growth is still possible, but
the field begins to attain systematic variations in the $z$ direction
which are more similar to those in a dynamo.
In that case, we would have to deal with a coupled system and a
direct comparison with the NEMPI growth rate would not be possible.
We return to this issue later in \Sec{HelicityProduction}.
}

\begin{figure}[h!]\begin{center}
\includegraphics[width=\columnwidth]{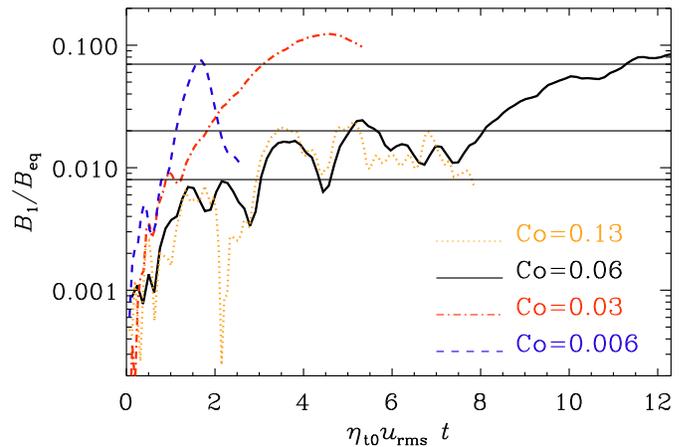}
\end{center}\caption[]{
Evolution of $B_1/\Beq$ for runs of which three are shown in \Fig{b005_th0}.
The three horizontal lines correspond to the approximate values of
$B_1/\Beq$ in the three rows of \Fig{b005_th0}.
}\label{pB1_vs_t_comp}\end{figure}

In \Fig{presults_Omdep} we compare with the MFS of LBKMR, who used
$\qpz=20$ and $\betap=0.167$ (corresponding to $\betastar=0.75$).
This set of parameters is based on a fit by \cite{KBKMR12c} for
$\kf/k_1=30$ and $\Rm=18$.
Note that the growth rates for the MFS are about 3 times larger than
those of the DNS.
As explained in the introduction, this might be caused by an inaccurate
estimate of the mean-field coefficients for these particular values
of $\kf/k_1$ and $\Rm$.
Indeed, according to Eq.~(22) of \cite{BKKR12}, who used $\kf/k_1=5$,
these parameters should be $\qpz=32$ and $\betap=0.058$
(corresponding to $\betastar=0.33$) for $\Rm=18$.
This assumes that these parameters are independent of the value of $\kf$,
which is not true either; see \cite{KBKMR12c}.
To clarify this question, we now perform 3-D MFS with this new set of
parameters.
Those results are also shown in \Fig{presults_Omdep}.
It turns out that with these parameters
the resulting growth rates are indeed much closer
to those of the DNS, suggesting that the former set of mean-field coefficients
might indeed have been inaccurate.
As alluded to in the introduction,
a reason for this might be the fact that for $\kf/k_1=30$ NEMPI is very
strong and leads to inhomogeneous magnetic fields for which the usual
determination of mean-field coefficients, as used by \cite{BKKR12},
is no longer valid, because for inhomogeneous magnetic fields
there would be additional terms in the expression for the
mean Reynolds stress \citep[cf.][]{KBKR13}.

\begin{figure*}[t!]\begin{center}
\includegraphics[width=\textwidth]{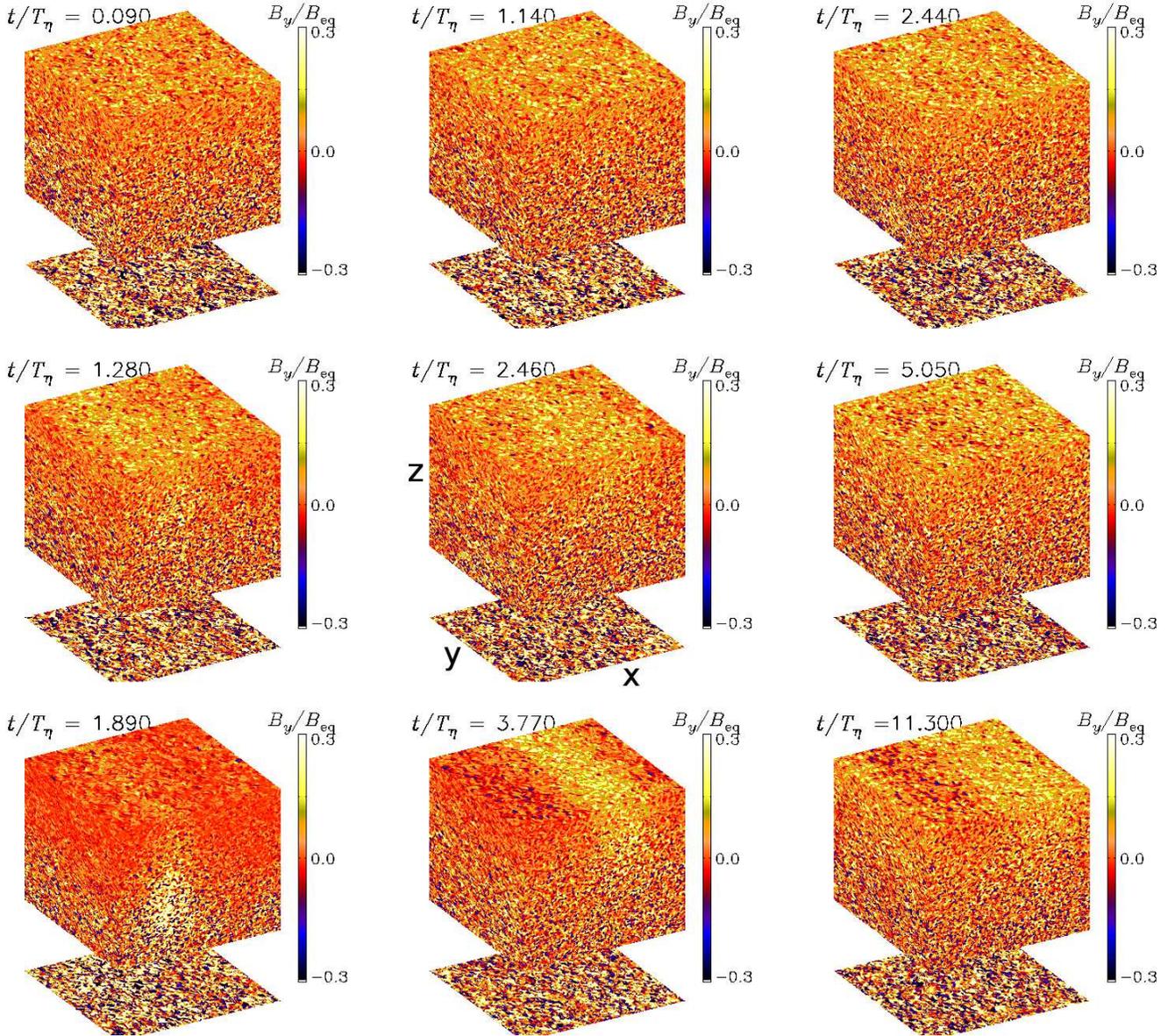}
\end{center}\caption[]{
$B_y$ at the periphery of the computational domain for $\Co=0.006$ (left),
0.03 (middle), and 0.06 (right) at the same times as in \Fig{b005_th0}.
The $x,y,z$ coordinates are indicated in the middle frame.
Note the strong surface effect for $\Co=0.03$ in the last time frame.
}\label{box005_th0}\end{figure*}

\begin{figure}[t!]\begin{center}
\includegraphics[width=\columnwidth]{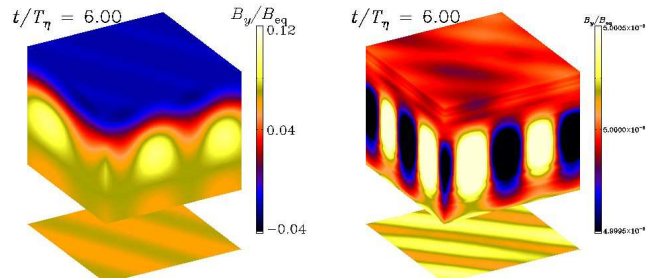}
\end{center}\caption[]{
Results of MFS of LBKMR showing $B_y$ at the periphery of the
computational domain for $\Co=+0.03$ in the LBKMR case (left) and with the new
set of parameters (right)
at the same time.
(The range in $B_y/B_0$ shown here is larger than that shown in LBKMR.)
}\label{12}\end{figure}

In \Fig{b005_th0} we show the $yt$-averaged $\meanB_y$ for $\Co=0.006$,
0.03, and 0.06 at different times.
When comparing results for different rotation rates, one should take
into account that the growth rates become strongly reduced.
Indeed, in the last row of \Fig{b005_th0} we have chosen
the times such that the amplitude of NEMPI is comparable for $\Co=0.006$
and 0.03, while for $\Co=0.06$ we have run much longer and the amplitude
of NEMPI is here even larger; see \Fig{pB1_vs_t_comp}, where we show
$B_1/\Beq$, which is the normalized magnetic field strength
for horizontal wavenumber $k=k_1$ in the top layers with $2\leq k_1z\leq3$.
It is clear that the formation of structures through NEMPI remains more
strongly confined to the upper-most layers as we increase the value of $\Co$.
Even for $\Co=0.13$ there is still noticeable growth of structures,
which is different from what is seen in MFS; see \Fig{presults_Omdep}.

\begin{figure*}[t!]\begin{center}
\includegraphics[width=\textwidth]{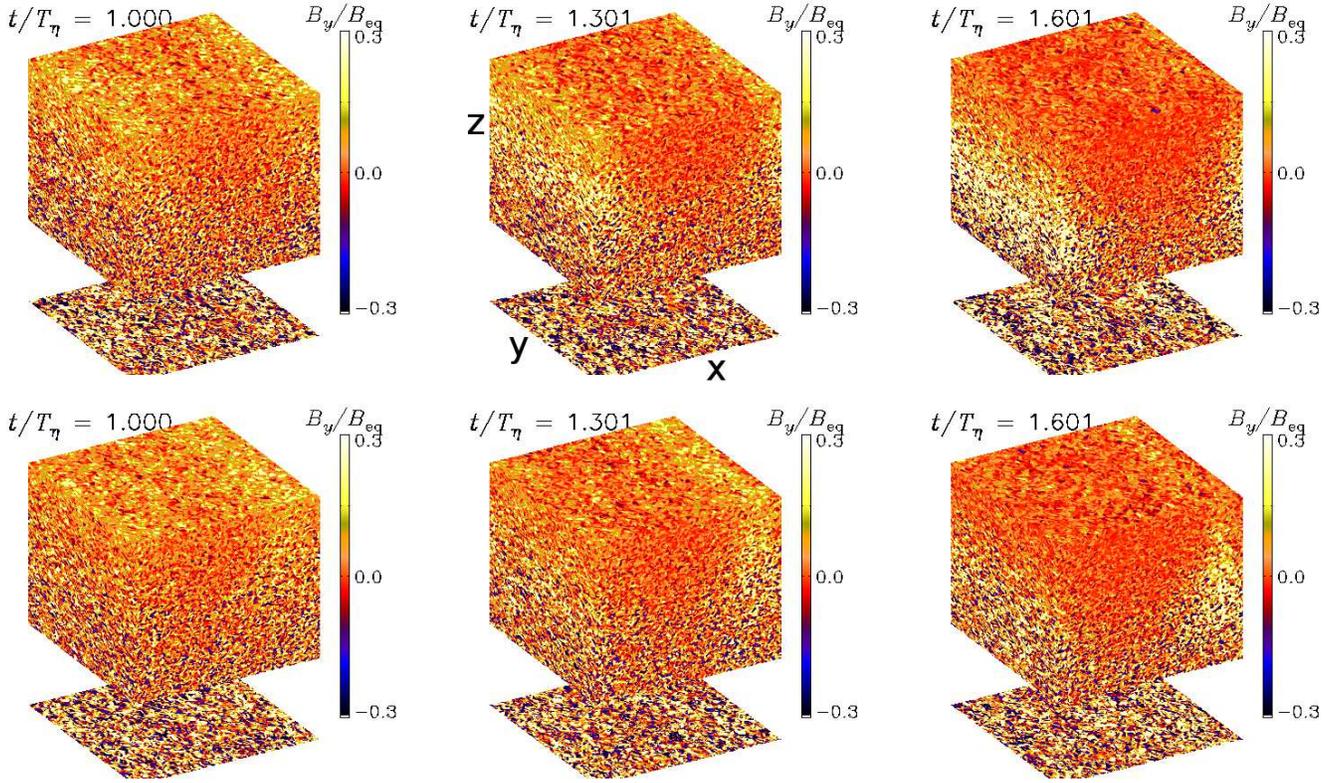}
\end{center}\caption[]{
$B_y$ at the periphery of the computational domain for $\Co=0.006$
and $\theta=45^\circ$ (upper row) and $90^\circ$ (lower row),
at three different times (from left to right).
The $x,y,z$ coordinates are indicated in the middle frame.
}\label{Om01_b005_th45_th90}\end{figure*}

These figures show the generation of structures that begin to sink
subsequently.
However, for $\Co=0.03$ and larger, this sinking is much less prominent.
Instead, the structures remain confined to the surface layers, which
is seen more clearly in visualizations of $B_y$ at the periphery of the
computational domain for $\Co=0.03$; see \Fig{box005_th0},
which is for approximately the same times as \Fig{b005_th0}.

To our surprise, the large-scale structures remain still independent
of the $y$-direction, which is clearly at variance with results of the
corresponding MFS.
In \Fig{12} we reproduce a result similar to that of LBKMR for $\Co=\pm0.03$.
Even at other angles such as $\theta=45^\circ$ and $90^\circ$,
no variation in the $y$-direction is seen; see \Fig{Om01_b005_th45_th90}.
The reason for this discrepancy between DNS and the corresponding MFS
is not yet understood.
Furthermore, the confinement of structures to the surface layers, which
is seen so clearly in DNS, seems to be absent in the corresponding MFS.

\subsection{Comparison of the 2-D and 3-D data of LBKMR}

\begin{figure}[h!]\begin{center}
\includegraphics[width=\columnwidth]{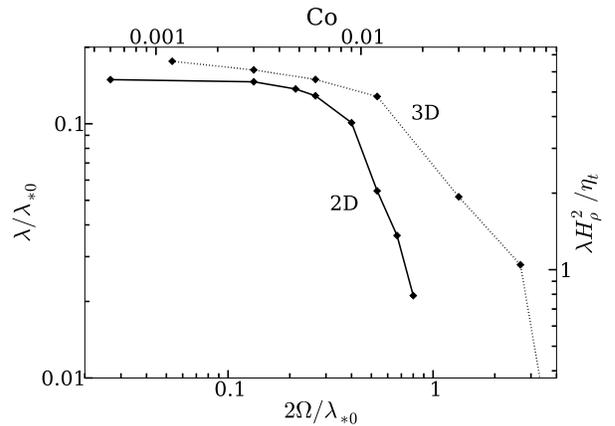}
\end{center}\caption[]{
Dependence of $\lambda/\lambda_{\ast0}$ on $2\Omega/\lambda_{\ast0}$
for $\theta=0^\circ$ in the 3-D and 2-D cases for $\theta=0$
(corresponding to the pole).
}\label{fig:MFSgrowth3Dvs2D}\end{figure}

The apparent lack of $y$ dependence of the large-scale magnetic field
in the DNS shows that this contribution to the magnetic field is
essentially two-dimensional.
In the lower panel of Figure~5 of LBKMR, a comparison between 2-D and 3-D
MFS was shown for $\Co\approx0.01$ as a function of latitude.
At the pole, the normalized growth rates were
$\lambda/\lambda_{\ast0}\approx0.07$ and 0.14
for 2-D and 3-D MFS, respectively.
This difference is smaller for smaller values of $\Co$, but it
increases with increasing values of $\Co$; see \Fig{fig:MFSgrowth3Dvs2D}.
\blue{
We note in passing that the 2-D result in this figure supersedes that of
Figure~3 of LBKMR, were $\lambda$ was determined from the amplification of
the total magnetic field (which includes the imposed field), rather than
the deviations of the magnetic field from the imposed one.
This resulted in a 4 times smaller estimate of $\lambda$.
Furthermore, the critical value of Co, above which NEMPI shuts off,
is now delayed by a factor of about 2--3.
}

The plot in \Fig{fig:MFSgrowth3Dvs2D} has been done for the
more optimistic set of mean-field parameters
($\qpz=20$ and $\betap=0.167$), but the essential
conclusions that the growth rates in 2-D and 3-D are similar
should not depend on this.
The remaining differences between DNS and MFS regarding the lack of
$y$ dependence of the mean field and the confinement of structures to
the surface layers might be related to absence of mean-field transport
coefficients other than $\qp$, $\etat$, and $\nut$.
By and large, however, the agreement between DNS and MFS
is remarkably good in that the predicted decline of NEMPI at rather
modest rotation rates is fully confirmed by DNS.

\section{Kinetic and magnetic helicity}

By adding rotation to our strongly stratified simulations,
we automatically also produce kinetic helicity.
In this section we quantify this, compare with earlier work, and
address the question whether this might lead to observable effects.
\blue{
All results presented in this section are based on time series with
error bars being estimated as the largest departure from any one third
of the full time series.
}

\subsection{Helicity production}
\label{HelicityProduction}

\begin{figure}[t!]\begin{center}
\includegraphics[width=\columnwidth]{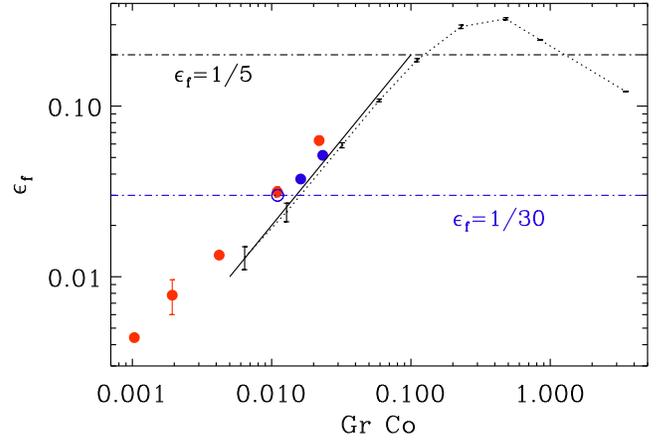}
\end{center}\caption[]{
Relative kinetic helicity spectrum as a function of $\Gr\,\Co$
\blue{
for $\Gr=0.03$ with $\Co=0.03$, 0.06, 0.13, 0.49,
and 0.66 (red and blue symbols)
}
compared with results from earlier simulations of \cite{BRK12} for $\Gr=0.16$
(small dots connected by a dotted line).
The solid line corresponds to $\epsf=2\Gr\,\Co$.
The two horizontal dash-dotted lines indicate the values of
$\epsf^\ast \equiv k_1/\kf$ above which
dynamo action is possible for $\kf/k_1=5$ and $30$, respectively.
\blue{
Runs without imposed field (blue filled symbols) demonstrate dynamo
action in two cases.
The blue open symbol denotes a case where the dynamo is close to marginal.
}
}\label{poum_Odep_g08}\end{figure}

In turbulence, the presence of rotation and stratification gives rise
to kinetic helicity and an $\alpha$ effect \citep{KR80,BGKKMR13}.
As a measure of kinetic helicity, we determine the normalized helicity
\EQ
\epsf=\overline{\oo\cdot\uu}/\kf\urms^2.
\EN
In \Fig{poum_Odep_g08} we compare
\blue{
our present runs at $\kf/k_1=30$ with those of \cite{BRK12}
}
at $\kf/k_1=5$ showing $\epsf$ versus $\Gr\,\Co$.
\blue{
For our present runs (red filled symbols), kinetic helicity is clearly
}
very small, which is a consequence of the small value of Co.
Compared with earlier runs at $\kf/k_1=5$, which gave
$\epsf\approx2\,\Gr\,\Co$, the present ones show
about twice as much helicity.
Interestingly, in the limit of rapid rotation the relative kinetic
helicity declines again when the product $\Gr\,\Co$ is larger than
about 0.5.
The maximum value of $\epsf$ that can be reached is about 0.3.
In a fully periodic domain, dynamo action would be possible when
$\epsf>(\kf/k_1)^{-1}$, which is $0.2$ in this case.
However, because of stratification and boundaries, the onset
is delayed and no dynamo action has been found in the simulations
of \cite{BRK12}.
\blue{
However, in the present case, dynamo action is possible for
$\epsf>1/30$ which leads to a Beltrami-like magnetic field
with variation in the $z$ direction.
Dynamo action is demonstrated in the absence of an imposed field,
which leads to slightly smaller values of $\epsf$ for the same value
of $\Gr\,\Co$ (see blue symbols in \Fig{poum_Odep_g08}).
The case $\Co=0.33$ is close to marginal and the field is slowly
decaying, which is in agreement with the expected position of the
marginal line.
}

\begin{figure}[t!]\begin{center}
\includegraphics[width=\columnwidth]{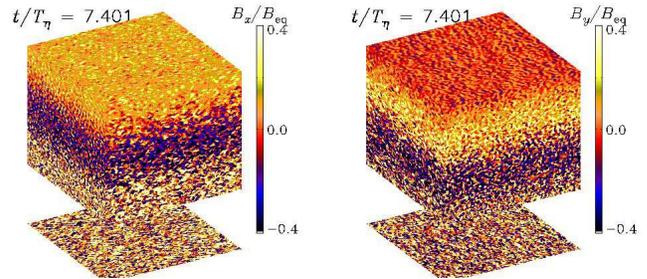}
\end{center}\caption[]{
Visualization of $B_x$ and $B_y$ for the run with $\Co=0.5$
showing dynamo action.
Note the clear signature of a Beltrami field showing variation
in the $z$ direction.
}\label{Om70_b0_th0}\end{figure}

\blue{
In \Fig{Om70_b0_th0} we show visualizations of $B_x$ and $B_y$
for a run with $\Co=0.5$ showing dynamo action.
Note the approximate phase shift of $90^\circ$ between $B_x$ and $B_y$
which has been seen in earlier simulations of $\alpha^2$-type dynamo
action from forced turbulence \citep{B01}.
As alluded to in \Sec{DNSandMFS}, the possibility of dynamo action might
be responsible for the continued growth found in DNS for $\Co\ge0.13$.
Visualizations of the $yt$-averaged $\meanB_y$ for $\Co=0.13$ and 0.31
show that structures with variation in the $x$ direction do still emerge
in front of a new component that varies strongly in the $z$ direction
and that becomes stronger as the value of $\Co$ is increased.
}

\blue{
Our results for $\Co=0.13$ and 0.31 might be examples
of a dynamo coupled to NEMPI.
Such coupled systems are expected to have an overall
enhancement of the growth.
This possibility, which is not included in the present mean-field model,
has recently been demonstrated in spherical geometry \citep{JBKMR13}
by coupling an $\alpha^2$ dynamo to NEMPI.
Looking at \Fig{presults_Omdep}, we are led to conclude that for
$\Co\ge0.13$, the coupled system with NEMPI and dynamo instability
is excited in a case where the dynamo alone would not be excited,
and that the growth rate begins to be larger than that of NEMPI alone.
Obviously, more work in that direction is necessary.
}

\begin{figure}[t!]\begin{center}
\includegraphics[width=\columnwidth]{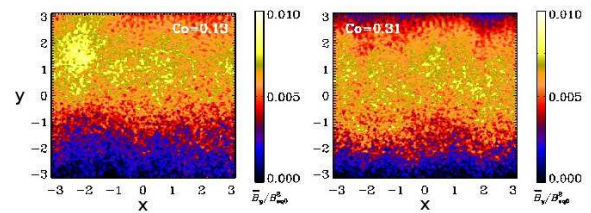}
\end{center}\caption[]{
Comparison of $yt$-averaged $\meanB_y$ for $\Co=0.13$ and 0.31.
}\label{Om20_50}\end{figure}

\subsection{Surface diagnostics}

As a consequence of the production of kinetic helicity,
the magnetic field should also be helical.
However, since magnetic helicity is conserved, and since it was zero initially,
it should remain zero -- at least on a dynamical time scale \citep{Ber84}.
This condition can be obeyed if the magnetic field is bi-helical, i.e.,
with opposite signs of magnetic helicity at large and small wavenumbers
\citep{See96,Ji99}.
We may now ask whether signatures of this could in principle
be detected at the solar surface.
To address this question, we use our simulation at intermediate
rotation speed with $\Co=0.03$, where magnetic flux concentrations
are well developed at the surface of the domain,
and compare with a larger value of 0.13.

\begin{figure}[t!]\begin{center}
\includegraphics[width=\columnwidth]{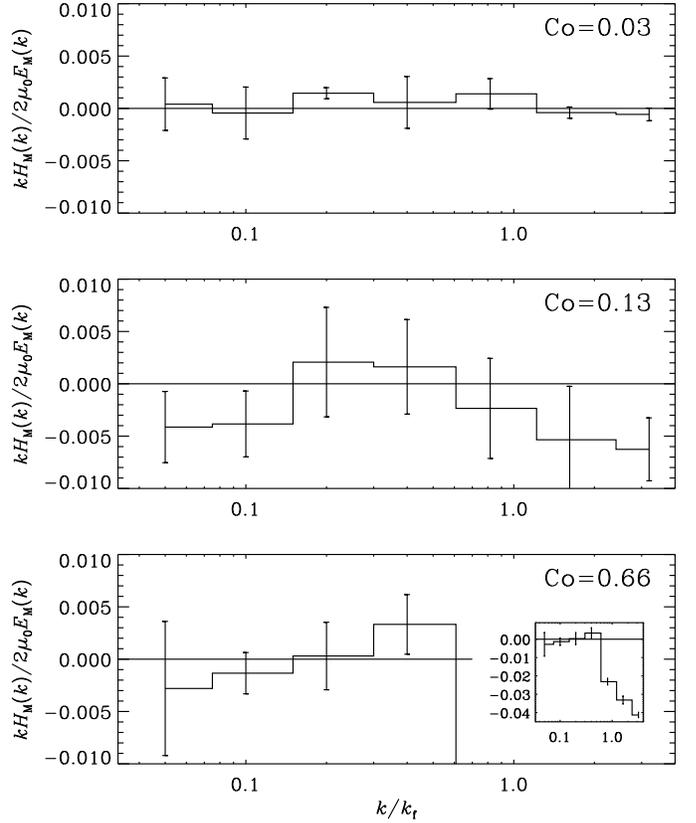}
\end{center}\caption[]{
\blue{
Normalized magnetic helicity
spectra for different values of the Coriolis number, $\Co$.
In all panels, the same range is shown, but for $\Co=0.66$ the
normalized helicity exceeds to this range and reaches $-0.05$.
}
}\label{pHHbinmm_EEbinmm_vs_kbin}\end{figure}

Measuring magnetic helicity is notoriously difficult, because it
involves the magnetic vector potential which is gauge dependent.
However, under the assumption of homogeneity and isotropy,
the Fourier transform of the magnetic correlation tensor is
\EQ
M_{ij}(\kk)=(\delta_{ij}-\hatk_i\hatk_j)\,
{\mu_0 E_{\rm M}(k)\over4\pi k^2}
-\epsilon_{ijk}{\ii k_kH_{\rm M}(k)\over8\pi k^2},
\label{IsotropicTensor}
\EN
where $\hatkk=\kk/k$ is the unit vector of $\kk$, and
$E_{\rm M}(k)$ and $H_{\rm M}(k)$ are magnetic energy and magnetic
helicity spectra, which obey the realizability condition,
$2\mu_0 E_{\rm M}(k)\ge k|H_{\rm M}(k)|$,
where the factor $2$ in front of $E_{\rm M}(k)$ is just a consequence
of the factor 1/2 in the definition of energy.
\cite{MGS82} used the solar wind data from {\it Voyager II} to determine
$H_{\rm M}(k)$ from the {\it in situ} measurements of $\BB$ and
\cite{BSBG11} applied it to measuring $H_{\rm M}(k)$ at high heliographic
latitudes where $H_{\rm M}(k)$ is finite and turned out to be bi-helical.
We now adopt the same method using Fourier transforms in the $y$ direction.
In the Sun, this corresponds to measuring $\BB$ along a $2\pi$ ring at
fixed polar latitude, where one might have a chance in observing
the full circumference at the same time.
In \Fig{pHHbinmm_EEbinmm_vs_kbin} we show the result for 
three values of Co.

It turns out that $H_{\rm M}(k)$ is compatible with zero for
our intermediate value of $\Co$.
For faster rotation ($\Co=0.13$), $H_{\rm M}(k)$ is negative both at large
wavenumbers ($k\gg\kf$), and positive (but still compatible with zero
within error bars) at intermediate wavenumbers ($0.15<k/\kf<0.6$).
For $k/\kf<0.15$, the magnetic helicity is again negative.
However, the error bars are large and rotation is already so fast
that structure formation via NEMPI is impossible.
It is therefore unclear whether meaningful conclusions can be drawn
from our results.

In the northern hemisphere of the Sun, a bi-helical spectrum
is expected where magnetic helicity is negative on all
scales except the largest ones where the $\alpha$ effect operates.
In this connection we remind the reader that this sense is reversed
in the solar wind far from the Sun \citep{BSBG11}.
This has also been seen in simulations of magnetic ejecta from a
dynamo-active sphere \citep{WBM11}, which may be explained by a
diffusive magnetic helicity transport \citep{WBM12}.

\subsection{Energy and helicity and velocity and magnetic fields}

\begin{figure}[t!]\begin{center}
\includegraphics[width=\columnwidth]{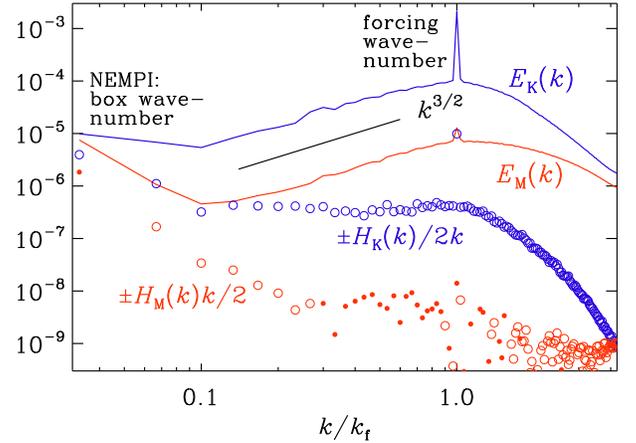}
\end{center}\caption[]{
Kinetic and magnetic energy and helicity spectra computed from the
full three-dimensional data set for $\Co=0.03$.
Positive (negative) values of spectral helicity are indicated
with filled (open) symbols.
Note the enhancement of spectral power at the smallest wavenumber
of the domain, $k_1$.
}\label{pspec_Om05_b005_th0_cont}\end{figure}

\blue{
To put the above considerations in relation to the actual helicity
content, we now compare with the magnetic and kinetic energy and
helicity spectra computed from the fully three-dimensional data set;
see \Fig{pspec_Om05_b005_th0_cont}.
The magnetic and kinetic helicity spectra are normalized by
$k/2$ and $1/2k$, respectively, so that one can estimate how
much the absolute values of these spectra are below their maximum possible
values given by the corresponding realizability conditions,
$|H_{\rm M}|k/2\le E_{\rm M}$ and $|H_{\rm K}|/2k\le E_{\rm K}$,
respectively.
}

\blue{
The spectra show that only at the largest scale the velocity and
magnetic fields have significant helicity, but they remain clearly
below their maximum possible values.
At large scales (small $k$), both helicities are negative (indicated by open
symbols), but the magnetic helicity is predominantly positive
at wavenumbers slightly below $\kf$.
This is consistent with \Fig{pHHbinmm_EEbinmm_vs_kbin}, which also
shows positive values, although only in the case of faster rotation ($\Co=0.13$).
Below the forcing scale, both energy spectra show a $k^{3/2}$ spectrum,
which is shallower than the white noise spectrum ($k^{2}$) and similar
to what has been seen in helically driven dynamos \citep{B01}.
Note also the uprise of magnetic and kinetic power at the smallest
wavenumber ($k=k_1$), which is again similar to helically driven dynamos,
but it is here not as strong as in the dynamo case.
}

\section{Conclusions and discussion}

The present work has confirmed the rather stringent restrictions of
LBKMR showing that NEMPI is suppressed already for rather weak rotation
($\Co\ga 0.03$).
\blue{
This demonstrates the predictive power of those earlier mean-field
simulations (MFS).
On the other hand, it also shows that the consideration of the
mere existence of a negative pressure is not sufficient.
We knew already that sufficiently strong stratification and
scale separation are two important necessary conditions.
In this sense, the existence of NEMPI might be a more fragile
phenomenon that the existence of a negative magnetic pressure,
which is rather robust and can be verified even in absence of
stratification \citep{BKR10}.
For the rather small Coriolis numbers considered here, no measurable
change of $\qp$ was seen in the simulations, which is in agreement with
our theory which predicts that the change is of the order $\Co^2$.
}

\blue{
Applied to the Sun with $\Omega=2\times10^{-6}\s^{-1}$, the strong
sensitivity of the instability to weak rotation implies
}
that NEMPI can only play a role in the upper-most layers where the
correlation time is shorter than $\Co/2\Omega\approx2$\,hours.
Although this value might change with changing degree of stratification,
it is surprising as it would exclude even the lower parts
of the supergranulation layer where $\tau$ is of the order of 1 day.
On the other hand, we have to keep in mind that our conclusions
based on isothermal models, should be taken with care.
It would therefore be useful to extend the present studies to polytropic
layers where the scale height varies with depth.
It should also be noted that weak rotation ($\Co=0.03$) enhances
the surface appearance.
At the same time, as we have argued in \Sec{DNSandMFS}, the sinking of
structures becomes less prominent, which suggests that they might
remain confined to the surface layers.
However, preliminary MFS do not indicate a significant
dependence of the eigenfunction on $\Co$ for values below 0.1.
Our interpretation, if correct, would therefore need to be
a result of nonlinearity.

If we were to apply NEMPI to the formation of active regions in the Sun,
we should keep in mind that the scale of structures would be 6--8 pressure
scale heights \citep{KBKMR12c}.
At the depth where the turnover time is about 2 hours, we estimate the
rms velocity to be about 500\,m/s, so the scale height would be about
3\,Mm, corresponding to a NEMPI scale of at least 20\,Mm.
This might still be of interest for explaining plage regions in the Sun.
Clearly, more work using realistic models would be required for making
more conclusive statements.

Regarding the production of kinetic helicity and the possible detection
of a magnetic helicity spectrum, our results suggest that the relative
magnetic helicity cannot be expected to be more than about 0.01.
This is a consequence of correspondingly low values of kinetic helicity.
We find that the normalized kinetic helicity is given by
$\epsf\approx2\Gr\,\Co$.
For the Sun, we expect $\Gr=(\kf H_\rho)^{-1}\approx0.16$,
which agrees with what is used in our simulations, leaving therefore
not much room for more optimistic estimates.
In this connection we should note that in
\cite{KBKMR12c} the value of $\kf H_\rho$ ($=\Gr^{-1}$) was
estimated based on stellar mixing length theory, using
$\ell_{\rm mix}=\alpha_{\rm mix}H_p$ for the mixing length
with $\alpha_{\rm mix}\approx1.6$ being an empirical parameter.
For isentropic stratification, the pressure scale height $H_p$ is
related to $H_\rho$ via $\gamma H_p\approx H_\rho$.
With $\kf=2\pi/\ell_{\rm mix}$ we obtain
$\kf H_\rho=2\pi\,\gamma/\alpha_{\rm mix}\approx2\pi$,
so $\Gr=(\kf H_\rho)^{-1}\approx0.16$.
We note here that, owing to a mistake,
we underestimated the value of
$\kf H_\rho$ by a factor of 2.6.
This factor also has an enhancing effect on the growth rate of NEMPI.
The correct value should then be larger
and would now be clearly faster than the turbulent--diffusive rate.
\blue{
Furthermore, as we have shown here, at the point where NEMPI begins
to be suppressed by rotation, effects related to dynamo action
reinforce the concentration of flux, even though the dynamo alone
would not yet be excited.
In this sense, the stringent restrictions of LBKMR from MFS
appear now less stringent in DNS.
It might be hoped that this new feature can eventually be
reproduced by MFS such as those of \cite{JBKMR13} that take
the $\alpha$ effect into account.
}

\begin{acknowledgements}
We thank Koen Kemel for helpful comments concerning the influence of
magnetic structure formation on the measurement of $\qpz$ and $\betap$
in DNS and an anonymous referee for useful suggestions that have led
to improvements in the presentation and a more thorough analysis.
Illa R.\ Losada was supported by PhD Grant `Beca de Investigaci\'on
CajaCanarias para Postgradua­dos 2011'.
This work was supported in part by
the European Research Council under the
AstroDyn Research Project No.\ 227952,
by the Swedish Research Council under the project grants
621-2011-5076 and 2012-5797 (IRL, AB),
by EU COST Action MP0806,
by the European Research Council under the Atmospheric Research Project No.\
227915, and by a grant from the Government of the Russian Federation under
contract No. 11.G34.31.0048 (NK, IR).
We acknowledge the allocation of computing resources provided by the
Swedish National Allocations Committee at the Center for
Parallel Computers at the Royal Institute of Technology in
Stockholm and the National Supercomputer Centers in Link\"oping, the High
Performance Computing Center North in Ume\aa and the Nordic High Performance
Computing Center in Reykjavik.
\end{acknowledgements}

\appendix

\section{The identities used in Sect.~\ref{theory}
for the integration in ${\bm k}$--space}

To integrate over the angles in ${\bm k}$--space in Sect.~\ref{theory}
we used the following identities \citep{RK04,RK07}:
\begin{eqnarray}
&&\bar K_{ij} = \int {k_{ij} \sin \theta \over 1 + a \cos^{2}
\theta} \,d\theta \,d\varphi = \bar A_{1} \delta_{ij} + \bar A_{2}
\beta_{ij} ,
\label{C22} \\
&& \bar K_{ijmn} = \int {k_{ijmn} \sin \theta \over 1 + a \cos^{2}
\theta} \,d\theta \,d\varphi = \bar C_{1} (\delta_{ij} \delta_{mn}
+ \delta_{im} \delta_{jn}
\nonumber \\
&& \quad+ \delta_{in} \delta_{jm}) + \bar C_{2}
\beta_{ijmn} + \bar C_{3} (\delta_{ij} \beta_{mn} + \delta_{im}
\beta_{jn} + \delta_{in} \beta_{jm}
\nonumber \\
&&\quad + \delta_{jm} \beta_{in} + \delta_{jn}
\beta_{im} + \delta_{mn} \beta_{ij}) ,
\label{C24}\\
&&\bar H_{ijmn}(a) = \int {k_{ijmn} \sin \theta \over (1 + a
\cos^{2} \theta)^{2} } \,d\theta \,d\varphi
\nonumber \\
&&\quad= - \biggl( {\partial
\over \partial b } \int {k_{ijmn} \sin \theta \over b + a \cos^{2}
\theta} \,d\theta \,d\varphi \biggr)_{b=1}
\nonumber \\
&&\quad= \bar K_{ijmn}(a) + a
{\partial \over \partial a} \bar K_{ijmn}(a) \;,
\label{C23}
\end{eqnarray}
where $\beta = \overline{B} / B_{\rm eq}$,
$ \hat\beta_{i} = \beta_{i} / \beta$, $\beta_{ij} = \hat \beta_{i}
\hat \beta_{j}$, and
\begin{eqnarray}
\bar A_{1} &=& {2 \pi \over a} \biggl[(a + 1) {\arctan (\sqrt{a}) \over
\sqrt{a}} - 1 \biggr] \;,
\nonumber\\
\bar A_{2} &=& - {2 \pi \over a} \biggl[(a + 3) {\arctan (\sqrt{a}) \over
\sqrt{a}} - 3 \biggr] \;,
\nonumber\\
\bar C_{1} &=& {\pi \over 2a^{2}} \biggl[(a + 1)^{2} {\arctan
(\sqrt{a}) \over \sqrt{a}} - {5 a \over 3} - 1 \biggr]  \;,
\nonumber\\
\bar C_{2} &=& \bar A_{2} - 7 \bar A_{1} + 35 \bar C_{1} \;,
\nonumber\\
\bar C_{3} &=& \bar A_{1} - 5 \bar C_{1} \; .
\label{P22}
\end{eqnarray}
In the case of $ a \ll 1 $ these functions are given by
\begin{eqnarray*}
\bar A_{1}(a) &\sim& {4 \pi \over 3} \biggl(1 - {1 \over 5} a
\biggr) \;, \quad \bar A_{2}(a) \sim - {8 \pi \over 15} a \;,
\\
\bar C_{1}(a) &\sim& {4 \pi \over 15} \biggl(1 - {1 \over 7} a
\biggr) .
\end{eqnarray*}
In the case of $ a \gg 1 $ these functions are given by
\begin{eqnarray*}
\bar A_{1}(a) &\sim& {\pi^{2} \over \sqrt{a}} - {4 \pi \over a}
\;, \quad \bar A_{2}(a) \sim - {\pi^{2} \over \sqrt{a}} + {8 \pi
\over a} \;,
\\
\bar C_{1}(a) &\sim& {\pi^{2} \over 4 \sqrt{a}} - {4 \pi \over 3
a} .
\end{eqnarray*}

The functions $A_{n}^{(m)}(\tilde \beta)$
are given by
\begin{eqnarray}
&& A_{n}^{(0)}(\tilde \beta) = {3 \tilde \beta^{2} \over \pi}
\int_{\tilde \beta}^{\tilde \beta{\rm Rm}^{1/4}}
{\bar A_{n}(X^{2}) \over X^{3}} \,d X  \;,
\label{X26}\\
&& A_{n}^{(2)}(\tilde \beta) = {3 \tilde \beta^{6} \over \pi}
\int_{\tilde \beta}^{\tilde \beta{\rm Rm}^{1/4}}
{\bar A_{n}(X^{2}) \over X^{7}} \,d X ,
\label{X27}\\
&& \int_0^1 \bar A_{n}(a(\bar \tau)) \bar \tau^m \, d\bar \tau = {2
\pi \over 3} A_{n}^{(m)}(\tilde \beta) ,
\label{X28}
\end{eqnarray}
and similarly for $C_{n}^{(m)}(\tilde \beta)$, where
$a = [\tilde \beta u_{0} k \tau(k) / 2]^{2}$,
$\tilde \beta = \sqrt{8} \; \overline{B} / B_{\rm eq}$,
and
$X^{2} = \tilde \beta^{2} (k/ k_{0})^{2/3} =\tilde \beta^{2}/\bar \tau = a$.
The explicit form of the functions
$A_{n}^{(m)}(\tilde \beta)$ and $C_{n}^{(m)}(\tilde \beta)$ for $m = 0; \, 2$
are given by
\begin{eqnarray}
A_{1}^{(0)}(\tilde \beta) &=& {1 \over 5} \biggl[2 + 2 {\arctan \tilde \beta
\over \tilde \beta^3} (3 + 5 \tilde \beta^{2}) - {6 \over \tilde \beta^{2}}  -
\tilde \beta^{2} \ln {\rm Rm}
\nonumber \\
& & - 2 \tilde \beta^{2} \ln \biggl({1 + \tilde \beta^{2} \over 1 + \tilde \beta^{2}
\sqrt{\rm Rm}}\biggr) \biggr] ,
\label{X40}\\
A_{2}^{(0)}(\tilde \beta) &=& {2 \over 5} \biggl[2 - {\arctan \tilde \beta \over
\tilde \beta^3} (9 + 5 \tilde \beta^{2}) + {9 \over \tilde \beta^{2}}
- \tilde \beta^{2} \ln
{\rm Rm}
\nonumber \\
& & - 2 \tilde \beta^{2} \ln \biggl({1 + \tilde \beta^{2} \over 1 + \tilde \beta^{2}
\sqrt{\rm Rm}}\biggr) \biggr] ,
\label{X41}\\
A_{1}^{(2)}(\tilde \beta) &=& {2 \over 63} \biggl[1 + 3 {\arctan \tilde \beta
\over \tilde \beta^3} (7 + 9 \tilde \beta^{2}) - {21 \over \tilde \beta^{2}} -
{3 \tilde \beta^{2} \over 2} M(\tilde \beta) \biggr] ,
\nonumber \\
\label{X42}\\
C_{1}^{(2)}(\tilde \beta) &=& {1 \over 33} \biggl[{2 \over 21}
+ {\arctan \tilde \beta
\over \tilde \beta^5} \biggl({99 \over 14}\tilde \beta^{4}
+ 11 \tilde \beta^{2} + {9 \over 2}\biggl)
\nonumber \\
& & - {19 \over 2\tilde \beta^{2}} - {9 \over 2\tilde \beta^{4}} -
{\tilde \beta^{2} \over 7} M(\tilde \beta) \biggr] ,
\end{eqnarray}
where $M(\tilde \beta) = 1 - 2 \tilde \beta^{2} + 2 \tilde \beta^{4}
\ln (1 + \tilde \beta^{-2})$.
Here we have taken into account that ${\rm Rm} \gg 1$.
For $\overline{B} \ll B_{\rm
eq} / 4 {\rm Rm}^{1/4} $ these functions are given by
\begin{eqnarray*}
A_{1}^{(0)}(\tilde \beta) &\sim& 2 - {1 \over 5} \tilde \beta^{2} \ln {\rm Rm},
\\
A_{2}^{(0)}(\tilde \beta) &\sim& - {2 \over 5} \tilde \beta^{2} \biggl[\ln {\rm
Rm} +  {2 \over 15}\biggr] ,
\\
A_{1}^{(2)}(\tilde \beta) &\sim& {2 \over 3}\biggl(1 - {3 \over 10}
\tilde \beta^{2}\biggr) , \quad A_{2}^{(2)}(\tilde \beta) \sim - {2 \over 5}
\tilde \beta^{2} ,
\\
C_{1}^{(2)}(\tilde \beta) &\sim& {2 \over 15}  \biggl(1 - {3 \over 14}
\tilde \beta^{2}\biggr) .
\end{eqnarray*}
For $B_{\rm eq} / 4 {\rm Rm}^{1/4} \ll  \overline{B} \ll B_{\rm eq} / 4$
these functions are given by
\begin{eqnarray*}
A_{1}^{(0)}(\tilde \beta) &\sim& 2 + {2 \over 5} \tilde \beta^{2} \biggl[2 \ln
\tilde \beta -  {16 \over 15} + {4 \over 7} \tilde \beta^{2} \biggr] ,
\\
A_{2}^{(0)}(\tilde \beta) &\sim& {2 \over 5} \tilde \beta^{2} \biggl[4 \ln \tilde \beta
-  {2 \over 15} - 3 \tilde \beta^{2} \biggr] .
\end{eqnarray*}
Other functions in this case have the same asymptotics as in the case of
$\overline{B} \ll B_{\rm eq} / 4 {\rm Rm}^{1/4} $.
For $  B \gg {B}_{\rm eq} / 4 $ these functions are given by
\begin{eqnarray*}
A_{1}^{(0)}(\tilde \beta) &\sim& {\pi \over \tilde \beta} - {3 \over \tilde \beta^{2}},
\quad A_{2}^{(0)}(\tilde \beta) \sim - {\pi \over \tilde \beta} + {6 \over
\tilde \beta^{2}} ,
\\
A_{1}^{(2)}(\tilde \beta) &\sim& {3 \pi \over 7 \tilde \beta} - {3 \over 2
\tilde \beta^2} , \quad A_{2}^{(2)}(\tilde \beta) \sim - {3 \pi \over 7 \tilde \beta}
+ {3 \over \tilde \beta^2} ,
\\
C_{1}^{(2)}(\tilde \beta) &\sim& {3 \pi \over 28 \tilde \beta} - {1 \over 2
\tilde \beta^2} .
\end{eqnarray*}

The functions $q_1(\beta)$ and $q_2(\beta)$ are given by
\begin{eqnarray}
&& q_1(\beta) = {1 \over 12} \Big[A_1^{(0)}(0) - A_1^{(0)}(4\beta)
- \half A_2^{(0)}(4\beta)
\nonumber\\
&& \quad - (\Omega \tau_0)^2 \Big(A_1^{(2)}(0)
 - 2C_1^{(2)}(0) - 10 A_1^{(2)}(4\beta)
\nonumber\\
&& \quad + 20 C_1^{(2)}(4\beta)
+ {9 \over 2 \pi} \left[\bar A_1(16\beta^2) -2 \bar C_1(16\beta^2)
\right] \Big) \Big],
 \label{B23}\\
&& q_2(\beta) = {1 \over 12} \Big[\half A_2^{(0)}(4\beta)
+ (\Omega \tau_0)^2 \Big(A_1^{(2)}(0)
 - 6C_1^{(2)}(0)
\nonumber\\
&& \quad - 10 A_1^{(2)}(4\beta) + 60 C_1^{(2)}(4\beta)
+ {9 \over 2 \pi} \big[\bar A_1(16\beta^2)
\nonumber\\
&& \quad -6 \bar C_1(16\beta^2) \big] \Big) \Big].
\label{B24}
\end{eqnarray}

%r e f
\newcommand{\ymonber}[3]{ #1, {Monats.\ Dt.\ Akad.\ Wiss.,} {#2}, #3}

%\vfill\bigskip\noindent\tiny\begin{verbatim}
%$Header: /var/cvs/brandenb/tex/illa/DNS_rotation/paper.tex,v 1.101 2013-04-05
%11:32:49 brandenb Exp $
%\end{verbatim}

\end{document}